\documentclass[twocolumn]{aastex631}
\usepackage{amsmath}
\begin{document}

\title{Strong H$_2$O and CO emission features in the spectrum of
KELT-20b driven by stellar UV irradiation}

\author[0000-0002-3263-2251]{Guangwei Fu}
\affiliation{Department of Astronomy, University of Maryland, College Park, MD 20742, USA; guangweifu@gmail.com}

\author[0000-0001-6050-7645]{David K. Sing}
\affiliation{Department of Physics and Astronomy, Johns Hopkins University, Baltimore, MD 21218, USA}

\author[0000-0003-3667-8633]{Joshua D. Lothringer}
\affiliation{Department of Physics, Utah Valley University, Orem, UT 84058, USA}

\author[0000-0001-5727-4094]{Drake Deming}
\author[0000-0003-2775-653X]{Jegug Ih}
\author[0000-0002-1337-9051]{Eliza Kempton}
\author[0000-0002-2110-6694]{Matej Malik}
\author[0000-0002-9258-5311]{Thaddeus D.\ Komacek}
\affiliation{Department of Astronomy, University of Maryland, College Park, MD 20742, USA; guangweifu@gmail.com}

\author[0000-0003-4241-7413]{Megan Mansfield}
\affiliation{Steward Observatory, University of Arizona, Tucson, AZ, USA}
\author[0000-0003-4733-6532]{Jacob L.\ Bean}
\affiliation{Department of Astronomy \& Astrophysics, University of Chicago, 5640 South Ellis Avenue, Chicago, IL 60637, USA}

\begin{abstract}
Know thy star, know thy planetary atmosphere. Every exoplanet with atmospheric measurements orbits around a star, and the stellar environment directly affects the planetary atmosphere. Here we present the emission spectrum of ultra-hot Jupiter KELT-20b which provides an observational link between host star properties and planet atmospheric thermal structure. It is currently the only planet with thermal emission measurements in the $T_{eq}\sim$2200K range that orbits around an early A-type star. By comparing it with other similar ultra-hot Jupiters around FGK stars, we can better understand how different host star types influence planetary atmospheres. The emission spectrum covers 0.6 to 4.5 $\mu m$ with data from TESS, HST WFC3/G141, and Spitzer 4.5 $\mu m$ channel. KELT-20b has a 1.4 $\mu m$ water feature strength metric of S$_{H_2O}$ = -0.097$\pm$0.02 and a blackbody brightness temperature difference of 528K between WFC3/G141 (T$_b$=2402$\pm$14K) and Spitzer 4.5 $\mu m$ channel (T$_b$=2930$\pm59$K). These very large H$_2$O and CO emission features combined with the A-type host star make KELT-20b a unique planet among other similar hot Jupiters. The abundant FUV, NUV, and optical radiation from its host star (T$_{eff}=8720\pm250$K) is expected to be the key that drives its strong thermal inversion and prominent emission features based on previous PHOENIX models calculations.

\end{abstract}
\nopagebreak

\section{Introduction}

Hot Jupiters (HJs) with atmospheric thermal inversions are expected to have spectral emission features. Almost all hot Jupiter emission spectra observed to date have shown muted \citep{mikal-evans_confirmation_2020} or non-existent 1.4 $\mu m$ water emission features \citep{parmentier_thermal_2018, mansfield_unique_2021, fu_hubble_2021-1, fu_hubble_2021}. Thermal dissociation of water and H- abundance \citep{arcangeli_h-_2018, parmentier_thermal_2018} are both expected to increase with temperature starting $\sim$2200K which will weaken the 1.4 $\mu m$ water emission feature. Indeed, despite detection of thermal inversions, water emission features have not been seen for planets with dayside temperatures above $\sim$2800K \citep{haynes_spectroscopic_2015, beatty_evidence_2017, stevenson_deciphering_2014, kreidberg_global_2018, arcangeli_h-_2018}. Host star spectral type has been predicted to be another determining factor of the HJ emission spectrum where increasing host star temperature strengthens planetary spectral emission features. This is due to absorption of the stronger FUV/UV flux from A-type stars by atomic metals and metal oxides (e.g., Fe I, Mg I, Ca I, TiO, VO, etc) which heats up the upper planetary atmospheric layers ($>$ 1 mbar) and drives stronger thermal inversions \citep{mansfield_unique_2021, lothringer_influence_2019, yan_temperature_2020}. Based on the combination of physical effects listed above, we expect planets orbiting A-type stars with low dayside temperatures to show the largest spectral emission features. The detection of large emission features on KELT-20b demonstrates that even planets with relatively low dayside temperatures can exhibit strong thermal inversions and emission features, driven by the abundant host star FUV/UV flux.

\section{Observations and data reduction}

The KELT-20b emission spectrum consists of observations from TESS, WFC3/G141, and Spitzer 4.5 $\mu m$ channel. The TESS dataset includes 17 eclipses from observations in sectors 14, 40, and 41. The WFC3/G141 dataset was collected on 2021-09-20 as part of the GO 16307 (PI: Guangwei Fu) program. The Spitzer 4.5 $\mu m$ channel eclipses come from archival data observed on 2019-02-21 as part of the GO 14059 (PI: Jacob Bean) program. All of the orbital parameters used for the TESS, WFC3, and Spitzer data reductions come from \cite{lund_kelt-20b_2017}.

\subsection{HST Analysis}

The WFC3/G141 eclipse dataset includes observations covering five consecutive HST orbits taken in spatial scan mode. Each frame was taken with the 512 $\times$ 512 pixel subarray in SPARS25 and NSAMP=5 setting. The forward scanning rate is 0.7 arcsec $s^{-1}$ and the exposure time is 69.6 seconds. 

A spatially scanned 2D spectrum is first extracted from each frame and cleaned to remove any hot pixels and cosmic rays. It is then summed vertically to obtain the 1D spectrum. Next, we normalize each 1D spectrum by its median flux and use the $scipy.interpolate.interp1d$ function to interpolate the 1D spectrum in the wavelength direction. The relative sub-pixel level horizontal shifts are then calculated based on the average of all spectra. Wavelength shifted corrected 1D spectra are then summed in the wavelength direction to form the white light eclipse lightcurve, which is fitted using \texttt{emcee} \citep{foreman-mackey_emcee_2013} with a combination of \texttt{BATMAN} \citep{kreidberg_batman_2015}, the \texttt{RECTE} charge trapping systematics model \citep{zhou_physical_2017}, HST orbital phase and a 2nd order polynomial of the wavelength shifts. Each wavelength channel eclipse lightcurve is then fit with the same routine but with a fixed best-fit whitelight eclipse time at a 0.5 orbital phase.

\subsection{TESS Analysis}

We collected 4, 7, and 6 eclipse visits of KELT-20b from the TESS sector 14, 40, and 41 datasets correspondingly. Each eclipse visit was cut out of the TESS lightcurve including 4 hours before and after the mid-eclipse point. Then we fit the cut-out eclipse lightcurve with a combination of a linear slope and the \texttt{BATMAN} model. The slope is then subtracted out of each visit and the 17 eclipses are then stacked together for the final fit with \texttt{BATMAN} to obtain the eclipse depth of 139$\pm$8 ppm. Our updated eclipse depth is consistent to within one sigma compared to \cite{wong_visible-light_2021} which obtained an eclipse depth of $111^{+35}_{-36}$ ppm using only the sector 14 data.

\subsection{Spitzer Analysis}

We analyzed archival Spitzer/IRAC data for two secondary eclipses of KELT-20b, obtained by program 14059 (J. Bean, P.I.).  These data are full phase curve observations, whereas we here concentrate only on the secondary eclipse.  We accordingly restricted our analysis to data covering orbital phases in a limited range centered on each eclipse. Those ranges were further constrained by the span of the data in a single Spitzer observational sequence.  We explored using slightly different phase limits and thereby verified that the derived eclipse depths are not sensitive to the temporal span of the out-of-eclipse baseline.  The data comprise 64-frame subarray cubes (32x32 pixels in each frame), having exposure times of 0.4-seconds per frame.  To extract photometry from these data, we used 11 different circular apertures, centered on the star using a 2-D Gaussian fit procedure, and subtracting the (minimal) median background intensity in each frame via a histogram fitting procedure.  

Our method of correcting for Spitzer's intra-pixel sensitivity variations used pixel-level-decorrelation (PLD, \citealp{deming_spitzer_2015}), as implemented by \citet{garhart_statistical_2020} (we used the same code).  The PLD fitting procedure bins the data in time and uses 12 pixels as basis vectors (a 4x4 square, minus the corner pixels).  Binning in time improves the precision of the pixel basis vectors, and it more effectively matches the red-noise character of the intra-pixel variations due to the IRAC instrument \citep{deming_spitzer_2015}.  We verified that the binning is not so extreme as to significantly alter the shape of the eclipse \citep{kipping_binning_2010}, and we avoid overfitting by requiring that the number of data points after binning remains at least 10 times greater than the number of fitted parameters.  The best bin size and photometric aperture radius are chosen by the code, based on minimizing the scatter in the Allan deviation relation \citep{allan_statistics_1966}, which expresses how the standard deviation of the residuals (data minus fit) varies as a function of bin size.  That relation is ideally an inverse square root, and minimizing the Allan deviation scatter provides an initial fit that is optimal over a large range of time scales sampled by the data.

\begin{figure*}
\centering
  \includegraphics[width=\textwidth,keepaspectratio]{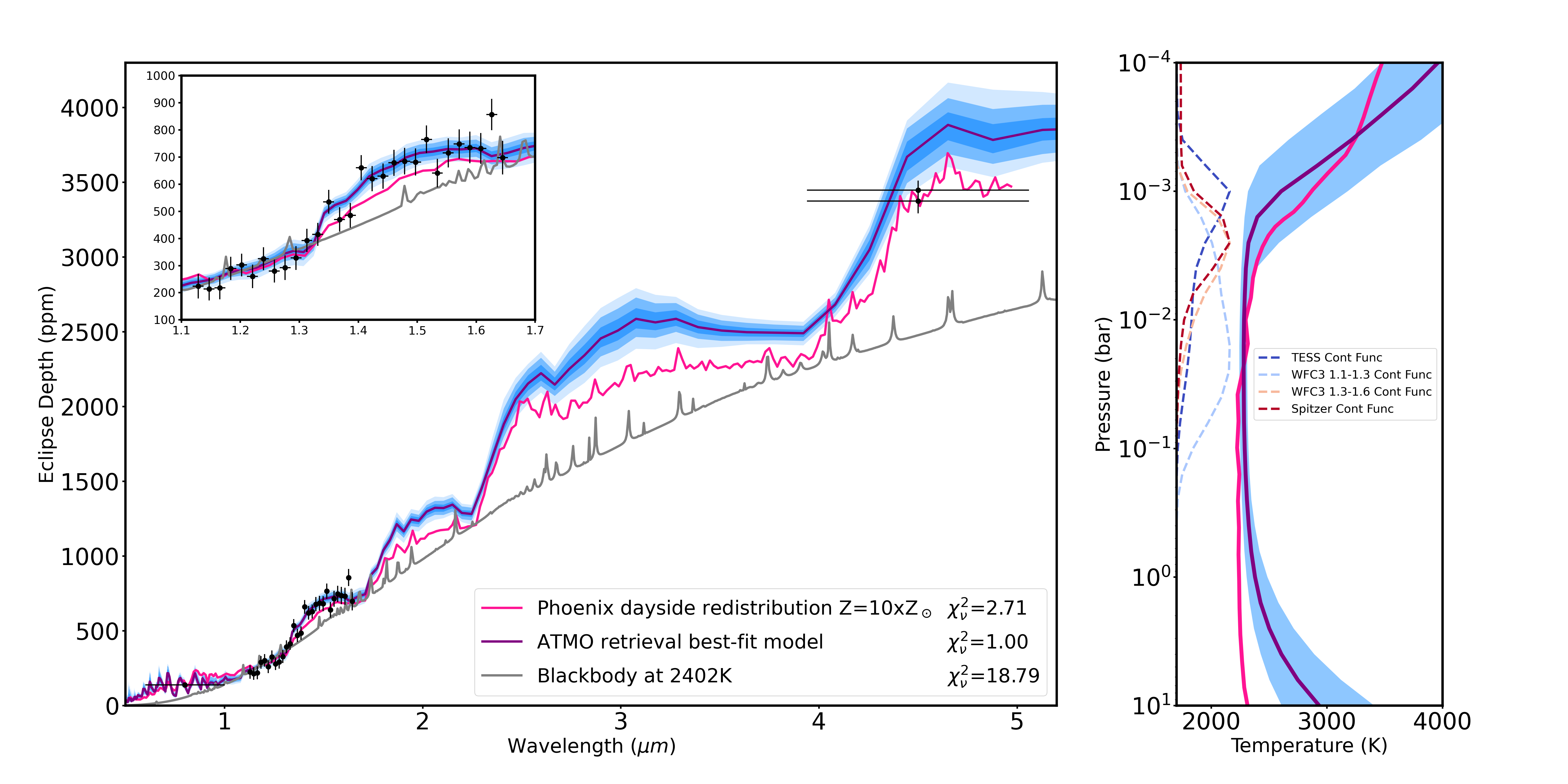}
  \caption{The emission spectrum of KELT-20b (black) is overplotted with the ATMO retrieval best-fit model (purple), PHOENIX forward model (pink), and blackbody model of 2402K (grey). The blue shaded regions represent 1 to 3 $\sigma$ uncertainties from ATMO retrieval. The TP profiles are shown on the right with corresponding colors for ATMO and PHOENIX best-fit models. The blue shaded region represents 1 $\sigma$ uncertainties. Contribution functions from ATMO for each wavelength channel are overplotted in dashed lines. The data probe pressure levels range from 100 to 1mbar with TESS being the lowest and the blue side of WFC3/G141 (1.1-1.3 $\mu m$) being the highest. We detected prominent H$_2$O and CO emission features indicating a strongly inverted TP profile as shown by both models. The excellent agreement of the TP profiles between ATMO and PHOENIX shows the dayside atmosphere is or very close to radiative equilibrium.}
  \label{fig:spectrum}
\end{figure*}

We use a quadratic baseline in time, fitting it simultaneously with the eclipse parameters and pixel coefficients.  Because KELT-20b is quite hot, we expect that even the limited span of data we are using out of eclipse will show intensity variations due to the phase curve.  Exploratory fitting indicated that the temporal baseline in these data is dominated by the phase curve of the planet, and not by a temporal ramp in instrumental sensitivity.  Accordingly, we forced the baseline to be flat during the eclipse (the planet being hidden then), and we verified that a quadratic is an adequate approximation to a sinusoid over the limited range of phase that we analyze.  

Our code is formally Bayesian, but our priors (e.g., for eclipse depth and central phase) are uniform, and we freeze the orbital parameters such as inclination and $a/R_s$ during the fit, as explained in Sec. 3.3 of \cite{garhart_statistical_2020}.  Under these conditions, the fitting process is equivalent to a $\chi$-squared minimization, but the process produces posterior distributions for the fitted parameters using a classic Markov Chain Monte Carlo (MCMC) chain with Gibbs sampling.  Our MCMC chains each comprise 800,000 samples, after a 10,000 sample burn-in.  Convergence is excellent, as verified by comparing the posterior distributions and Gelman-Rubin parameter for four independent chains at each eclipse.  Our adopted eclipse depths and error bars are based on a Gaussian fit to the posterior distributions.  The depths that we derive for the two independent eclipses are $3448\pm64$ and $3375\pm82$ ppm, in excellent mutual agreement.

\section{ATMO retrieval and PHOENIX model}

The full emission spectrum (Table \ref{eclipse_spectrum}) of KELT-20b from 0.6 to 4.5 $\mu m$ is presented in Figure \ref{fig:spectrum}. To interpret the spectrum, we performed ATMO retrieval \citep{amundsen_accuracy_2014, drummond_effects_2016, goyal_library_2018, tremblin_fingering_2015, tremblin_cloudless_2016} analysis and PHOENIX self-consistent forward model \citep{lothringer_extremely_2018} comparison. Having both retrieval and self-consistent atmospheric models allows us to cross-check the results to ensure more physically robust and consistent interpretations. The priors used for the ATMO retrieval are the following: log(Z/Z$_{\Sun}$)=-2 to 2; logg=2.5 to 3.46; log(K$_{IR}$)=-5 to -0.5; log($\gamma$/IR)=-4 to 2; $\beta$= 0 to 1.5; log(C/C$_{\Sun}$)=-2 to 2; log(O/O$_{\Sun}$)=-2 to 2. The best-fit retrieved spectrum (Figure \ref{fig:spectrum}) with a $\chi^2_{\nu}$ of 1.00 shows very prominent H$_2$O and CO emission features at 1.4 and 4.5 $\mu m$, respectively. The retrieved TP profile is highly inverted starting around 1mbar with a rapid increase of $\sim$1000K from $\sim$7 to 0.3mbar. The contribution functions of different wavelength channels of the emission spectrum show that the TESS band is probing the highest layers while the blue side (1.1-1.3 $\mu m$) of the WFC/G141 band is probing the deepest parts of the atmosphere. The red side (1.3-1.6 $\mu m$) of the WFC/G141 band and Spitzer 4.5 $\mu m$ have similar flux contributions from the 10 to 1mbar region where H$_2$O and CO have a relatively higher abundance (VMR $\sim$8$\times$10$^{-3}$). The retrieved metallicity, carbon, and oxygen (Figure \ref{fig:atmo_corner}) abundances are $\sim$3.9, 10 and 17 times higher than the solar values, but the retrieved C/O ratio of $0.454^{+0.211}_{-0.205}$ is consistent with the solar value to within one sigma. 

We also ran a set of self-consistent PHOENIX forward models assuming dayside heat redistribution, Local Thermodynamic Equilibrium (LTE), and solar C/O ratio with 0.1, 1, and 10 times solar metallicity. The best-fit model is 10 times metallicity with a $\chi^2_{\nu}$ of 2.71 showing similar strong H$_2$O and CO emission features compared to the retrieved ATMO best-fit spectrum. We consider this to be a very good forward model fit to the data considering there is only one parameter that was varied being the overall metallicity. The TP profile from PHOENIX is also in excellent agreement to within one sigma of retrieved ATMO TP profile among the pressure levels the data are probing. The matching TP profiles between retrieval and forward model indicate the atmosphere of KELT-20b around $\sim$10 - 1 mbar range is very close to LTE.

\begin{figure}
\centering
  \includegraphics[width=0.5\textwidth,keepaspectratio]{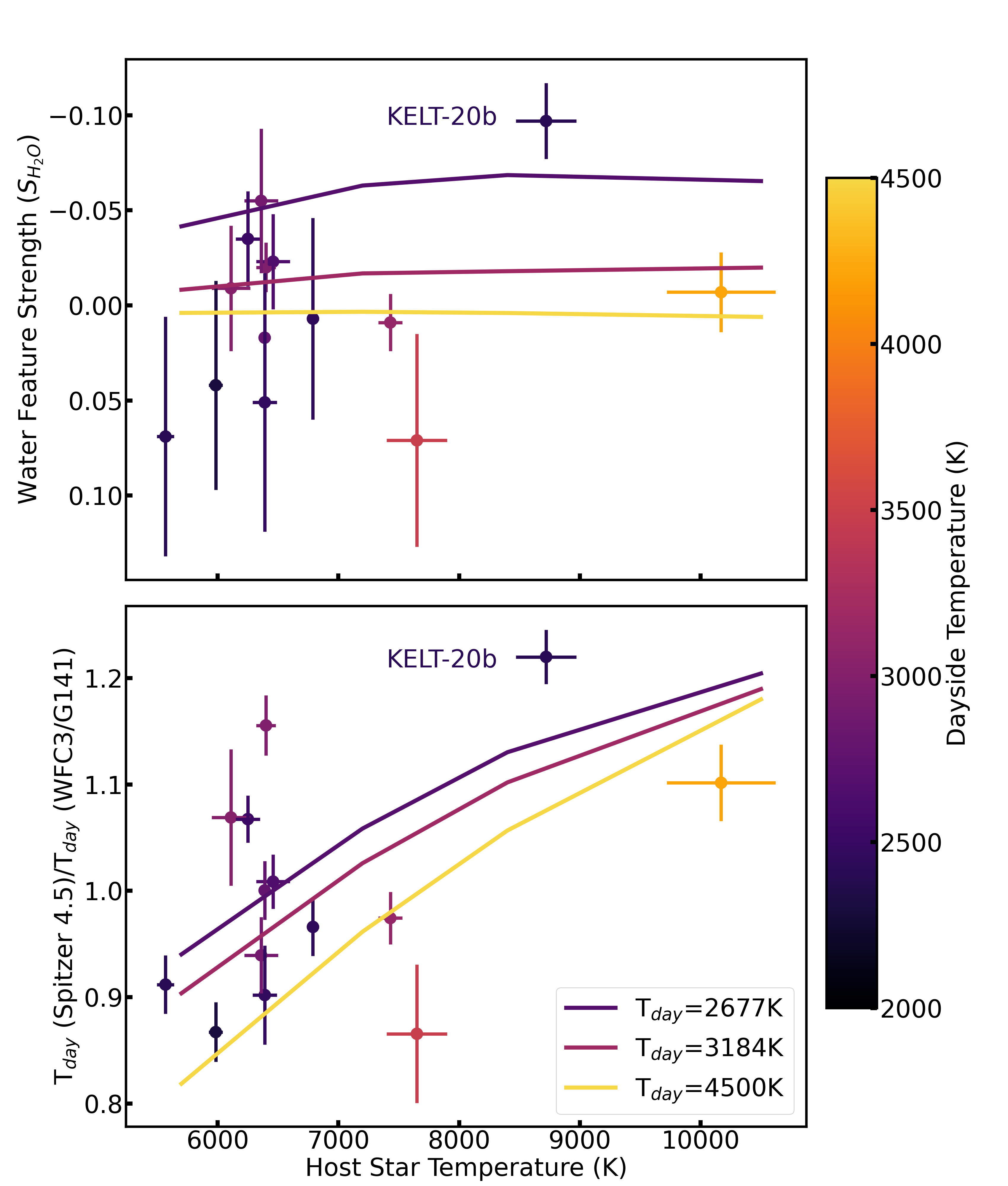}
  \caption{KELT-20b emission spectrum metrics compared with other hot Jupiters with dayside temperature higher than 2200K overplotted with PHOENIX models assuming solar metallicity and C/O ratio. The top panel shows the water feature strength metric S$_{H_2O}$ from \cite{mansfield_unique_2021} and KELT-20b is located in a unique parameter space of high host star temperature combined with low dayside temperature. The large water emission feature from KELT-20b is consistent with the PHOENIX model predictions. The bottom panel shows the Spitzer 4.5 $\mu m$ blackbody brightness temperature relative to the WFC3/G141 out of water band blackbody brightness temperature. The large difference between the two bands in KELT-20b indicates a prominent CO emission feature.}
  \label{fig:waterfeature}
\end{figure}

\section{Discussion}

The measured blackbody brightness temperatures are based on a PHOENIX stellar model \citep{husser_new_2013} grid (logg=4.5 and logZ=0) interpolated to T$_{eff}$ = 8720K are 2402$\pm$14K for the HST/WFC3 G141 band and 2930$\pm59$K for the Spitzer 4.5 $\mu m$ band. The prominent H$_2$O and CO emission features of KELT-20b make it unique compared to other UHJs within a similar equilibrium temperature range (Figure \ref{fig:waterfeature}) such as WASP-76b \citep{fu_hubble_2021-1}, WASP-121b \citep{evans_ultrahot_2017} and HAT-P-7b \citep{mansfield_hstwfc3_2018}. 

\subsection{S$_{H_2O}$ water feature strength}

KELT-20b has a 1.4 $\mu m$ water feature strength metric of S$_{H_2O}$ = -0.097$\pm$0.02 as defined in \cite{mansfield_unique_2021} which compares the 1.35 - 1.48 $\mu m$ ``in-band" part of the emission spectrum to the blackbody model fit based on the two ``out-of-band" regions (1.22 - 1.33 and 1.52 - 1.61 $\mu m$) of the spectrum. A positive value would suggest a water absorption feature and a negative value would indicate a water emission feature. In comparison with other hot Jupiters that have measured dayside temperatures exceeding 2200K, KELT-20b has a very large water emission feature (Figure \ref{fig:waterfeature} top panel). It is a unique planet with a relatively cool dayside but high host star temperature. High FUV/UV flux from the star drives a strong thermal inversion, while the low dayside temperature suppresses the thermal dissociation of water and the H- continuum opacity source. To understand the combined atmospheric effects from the planet and host star temperatures, we generated a grid of self-consistent PHOENIX models assuming solar metallicity and C/O ratio. For cooler planets, we see increased water emission feature amplitudes around hotter stars, but as planet temperature increases the water feature diminishes due to thermal dissociation and the raising of H- opacity. 

This combined effect of the planet and host star temperature on water emission feature strength has also been demonstrated with a different independent set of atmospheric models as shown in panel c of figure 4 in \cite{mansfield_unique_2021}. As host star temperature increases, the water emission feature is also expected to increase the most for hot Jupiters with dayside temperature between $\sim$2200 and 2500K and then taper as planets become even hotter.

\subsection{WFC3/G141 to Spitzer 4.5 $\mu m$ brightness temperature difference}

KELT-20b also has a large measured blackbody brightness temperature difference between ``out-of-band" WFC3/G141 (T$_b$=2402$\pm$14K) and the Spitzer 4.5 $\mu m$ (T$_b$=2930$\pm59$K) band which indicates a strong CO emission feature around 4.5 $\mu m$. The high relative blackbody brightness temperature ratio between the two bands ($Tb_{G141}/Tb_{4.5}$) from KELT-20b stands out compared to other hot Jupiters \citep{garhart_statistical_2020, baxter_evidence_2021} (Figure \ref{fig:waterfeature} bottom panel). In the absence of H- opacity, WFC3/G141 probes the H$_2$O and continuum while Spitzer 4.5 $\mu m$ band measures the CO spectral feature. These two bands can probe different pressure levels depending on the TP profile. In an inverted TP profile, Spitzer 4.5 $\mu m$ band usually probes higher up in the atmosphere as CO can exist at low-pressure levels with the high thermal dissociation temperature. However, as H- continuum opacity abundance raises with planet temperature, the photosphere pressure levels probed by these two bands converge. This is shown in the same PHOENIX model grid as described in the section above, as at a given host star temperature, increasing planet temperature decreases $Tb_{G141}/Tb_{4.5}$ as they both start to probe H- in the upper atmosphere. On the other hand, as the host star temperature increases, a higher amount of FUV/UV radiation gets absorbed in the upper atmosphere and drives a stronger thermal inversion which enhances the CO emission feature. At the same time, less flux can reach deeper down in the atmosphere which leads to a lower continuum temperature probed in the WFC3/G141 band. The $Tb_{G141}/Tb_{4.5}$ value decreases as planet temperature increases due to higher H- opacity raising the photosphere in the WFC3/G141 band and reducing the temperature difference measured between the two bands. Therefore we expect to see large $Tb_{G141}/Tb_{4.5}$ values around cooler planets with hotter host stars as shown in KELT-20b. This is consistent with the observational implications predicted by \cite{lothringer_influence_2019} (see their Figure 8) which suggested a larger relative emission flux difference between these WFC3/G141 and Spitzer 4.5 $\mu m$ bands as host star temperature increases. Since the PHOENIX model grid assumes solar metallicity and C/O ratio values, the data-model deviations also demonstrate the atmospheric composition diversity within the hot Jupiter population. 

\begin{table}
\begin{small}
\caption{\textbf{KELT-20b eclipse spectrum}}
\begin{tabular}{cccc}
\hline\hline & \\[-2ex]
\multicolumn{1}{c}{\shortstack{Wavelength \\ midpoint \\ ($\mu$m)}} & 
\multicolumn{1}{c}{\shortstack{Bin \\ width \\ ($\mu$m)}} & 
\multicolumn{1}{c}{\shortstack{Eclipse \\ Depth \\ (ppm)}} & 
\multicolumn{1}{c}{\shortstack{Uncertainty \\ (ppm)}}\\[1ex]
\hline 
\setlength\extrarowheight{3pt}
0.800	&	0.2000	&	139	&	8	\\
1.130	&	0.0092	&	224	&	45	\\
1.150	&	0.0092	&	214	&	42	\\
1.170	&	0.0092	&	218	&	43	\\
1.180	&	0.0092	&	289	&	43	\\
1.200	&	0.0092	&	302	&	42	\\
1.220	&	0.0092	&	261	&	44	\\
1.240	&	0.0092	&	325	&	43	\\
1.260	&	0.0092	&	280	&	42	\\
1.280	&	0.0092	&	292	&	45	\\
1.290	&	0.0092	&	328	&	44	\\
1.310	&	0.0092	&	393	&	43	\\
1.330	&	0.0092	&	415	&	42	\\
1.350	&	0.0092	&	535	&	44	\\
1.370	&	0.0092	&	470	&	45	\\
1.390	&	0.0092	&	485	&	46	\\
1.410	&	0.0092	&	660	&	47	\\
1.420	&	0.0092	&	621	&	47	\\
1.440	&	0.0092	&	630	&	46	\\
1.460	&	0.0092	&	679	&	48	\\
1.480	&	0.0092	&	685	&	50	\\
1.500	&	0.0092	&	682	&	50	\\
1.520	&	0.0092	&	765	&	51	\\
1.530	&	0.0092	&	641	&	53	\\
1.550	&	0.0092	&	715	&	54	\\
1.570	&	0.0092	&	748	&	54	\\
1.590	&	0.0092	&	735	&	58	\\
1.610	&	0.0092	&	732	&	57	\\
1.630	&	0.0092	&	857	&	58	\\
1.640	&	0.0092	&	698	&	62	\\
4.500	&	0.5600	&	3448	&	64	\\
4.500	&	0.5600	&	3375	&	82	\\
\hline 
\label{eclipse_spectrum}
\end{tabular}
\end{small}
\end{table}

\subsection{High-resolution spectroscopy}

KELT-20b has been observed during transit from the ground with multiple high-resolution spectroscopy facilities including HARPS-N, CARMENES, and EXPRES. Absorption features from neutral and ionized heavy metals including FeI, FeII, CaII, NaI, MgI, and Cr II have been detected with high confidence and confirmed from different independent studies \citep{nugroho_searching_2020, hoeijmakers_high-resolution_2020, casasayas-barris_atmospheric_2019, stangret_detection_2020}. The numerous detections of heavy metal species in the upper atmosphere layers through high-resolution transmission spectroscopy are consistent with a strongly inverted TP profile. These metal atoms are likely responsible for the thermal inversion by absorbing the abundant FUV, NUV, and optical flux from the star and then heating up the upper layers. The strong emission features of KELT-20b and the bright host star make it an ideal target for follow-up dayside high-resolution spectroscopy which can access the CO and H$_2$O features beyond 2 $\mu m$ to further constrain the metallicity, C/O ratio, and TP profile \citep{line_solar_2021}.

\section{Conclusion}

We present the emission spectrum of the ultra-hot Jupiter KELT-20b from 0.6 to 4.5 $\mu m$ showing strong H$_2$O and CO emission features. The H$_2$O feature strength is calculated through the S$_{H_2O}$ index and the CO feature is inferred by the differential brightness temperature measured between the WFC3/G141 and Spitzer 4.5 $\mu m$ channel bands. KELT-20b stands out among all other similar hot Jupiters with strong emission features while other UHJs have mostly shown featureless blackbody-like emission spectra. Our results imply that the unique early A-type host star of KELT-20b is the key difference that drives its stronger thermal inversion compared to other UHJs as predicted by \cite{lothringer_influence_2019}. KELT-20b provides direct observational evidence linking host star properties to planetary thermal structure, which adds host star property as the new parameter space to explore for our understanding of exoplanet atmospheres.

KELT-20b is the coolest planet among the only few UHJs (WASP-33b, Kepler-13Ab, and KELT-9b) that have been characterized around early A-type stars. The large differences in equilibrium temperatures within this very small sample size make it infeasible for a statistically significant comparison study. This calls for more follow-up atmospheric characterization of hot Jupiters that orbit A stars. HST will still be valuable in probing the water feature in the near-infrared for targets with favorable signal to noise. However, with Spitzer decommissioned, JWST will be the ideal telescope to access the H$_2$O and CO emission features in the infrared.

\begin{figure*}
\centering
  \includegraphics[width=\textwidth,keepaspectratio]{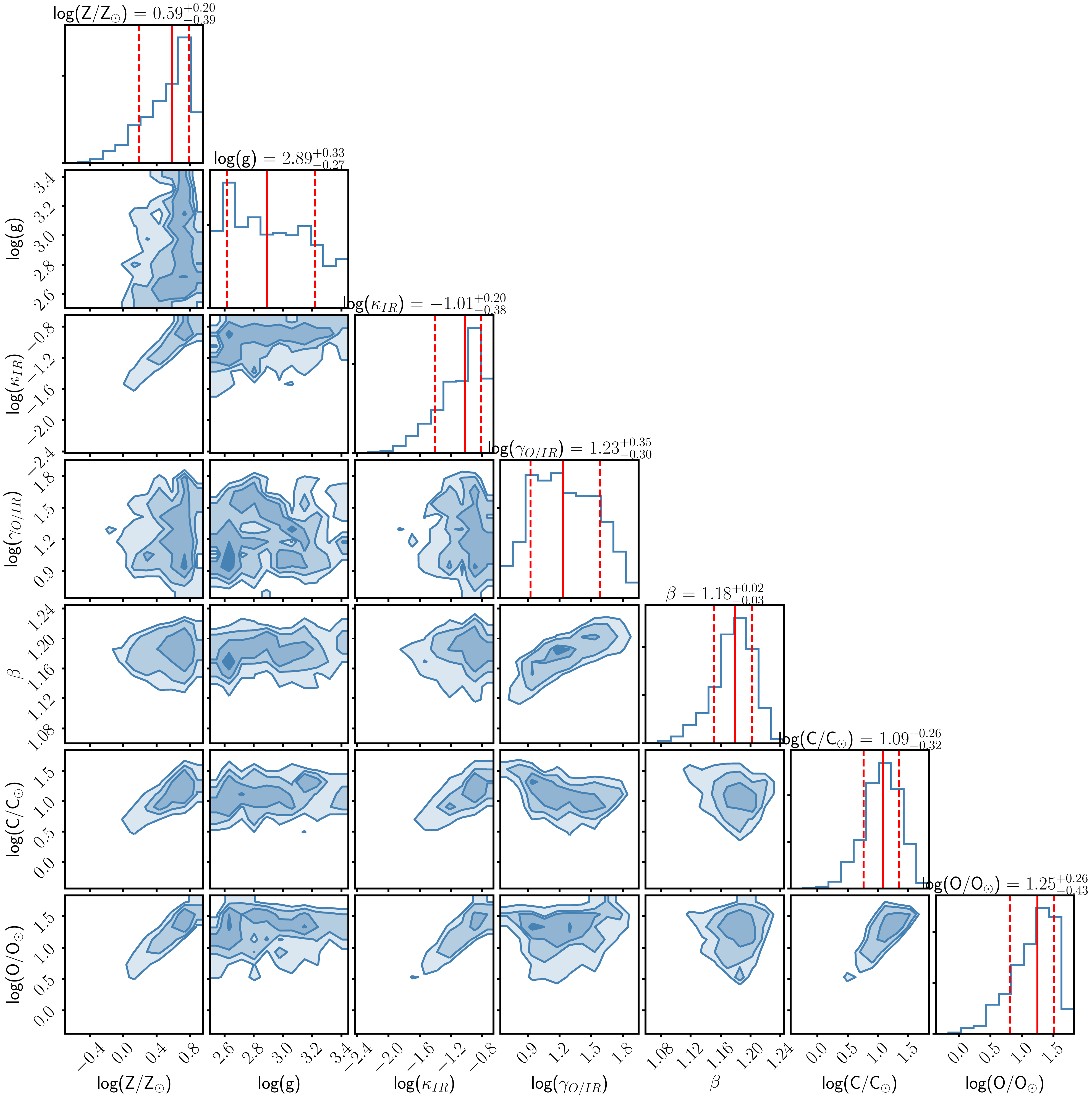}
  \caption{Posterior distribution of ATMO retrieval. The retrieved metallicity, carbon, and oxygen abundances are $\sim$3.9, 10, and 17 times higher than the solar values, but the retrieved C/O ratio of $0.454^{+0.211}_{-0.205}$ is consistent with the solar value to within one sigma. The log(g) is not well constrained as the emission spectrum is not sensitive to surface gravity.}
  \label{fig:atmo_corner}
\end{figure*}

\clearpage

\bibliography{references}

\end{document}